# A Hybrid DOS-Tolerant PKC-Based Key Management System for WSNs

Hamzeh Ghasemzadeh[1*], Ali Payandeh [2], Mohammad Reza Aref [3]

[1] Electrical Engineering, Azad University of Damavand, Tehran, Iran
[2] ICT Departments, Malek-e-Ashtar University of Technology, Tehran, Iran
[3] Electrical Engineering, Sharif University of Technology, Tehran, Iran
[*] hamzeh_g62@yahoo.com

**Abstract:** Security is a critical and vital task in wireless sensor networks (WSNs), therefore different key management systems have been proposed, many of which are based on symmetric cryptography. Such systems are very energy efficient, but they lack some other desirable characteristics. On the other hand, systems based on public key cryptography (PKC) have those desirable characteristics, but they consume more energy. Recently based on authenticated messages from base station (BS) a new PKC–based key agreement protocol was proposed. We show this method is susceptible to a form of denial of service (DOS) attack where resources of the network can be exhausted with bogus messages. Then, we propose two different improvements to solve this vulnerability. Simulation results show that these new protocols retain desirable characteristics of the basic method and solve its deficiencies.

## 1. Introduction

WSNs have attracted researchers from various fields over the past decade. These specialized networks are decentralized, self-organized and can be deployed without requiring the existence of a supporting infrastructure. Basically, they serve as an interface to the real world and gather some physical information from their surroundings. Thus, they have found a wide range of applications. Romer et al. surveyed many practical WSN projects [1]. Other more recent applications of WSNs include mining underground coal [2], environmental disaster monitoring [3], monitoring soccer players for injuries [4], laboratory tutoring [5], secure capturing of voice [6], military vehicle tracking [7], and many more. Unfortunately wireless connectivity, absence of physical protection, and the unattended deployment make WSNs prone to different types of attack. Consequently, for gathering reliable information these networks should be protected with appropriate security mechanisms. Many of which rely on existence of secure keys between different nodes of the network, a task that key management system (KMS) addresses. Recently, many KMS have been proposed for WSNs.

First, a method based on the probabilistic pre-distribution of subsets of a key-pool was proposed [8]. This method had low resiliency and its connectivity was poor. Later, methods based on symmetric polynomials and generating matrices of linear codes were proposed [9, 10]. These methods solved problem of connectivity, but they had threshold resiliency and addition of new nodes to the network was hard. LEAP was designed to support secure in-network processing [11]. But if the transitory initial key of LEAP is discovered, security of the entire network is compromised. In [12] a hash-based mechanism is employed to enhance the resiliency of key pre-distribution schemes against node capture. It was shown that this scheme improves resiliency of q-composite method [13]. But this scheme cannot achieve perfect resiliency and it inherits other undesirable characteristics of the basic method. Çamtepe et al. proposed a key pre-distribution scheme based on symmetric balanced incomplete block design [14]. Their scheme has good connectivity but it does not scale very well. To improve scalability of pre-distribution based systems, mapping from unitals to key pre-distribution was proposed [15]. But it does not guarantee perfect key sharing. Finally, biometric-based authentication and a two factor authentication based on attribute and password were proposed in [16, 17], respectively.

Another possible path to KMS is to use public key cryptography (PKC). PKC-based systems have many desirable characteristics. They provide perfect resiliency and perfect global, local, and node connectivity [18]. Furthermore, they are scalable and extensible [18]. On the other hand their energy consumption is high. In the past decade different PKC-based KMSs have been proposed. First, feasibility of PKC-based KMS in WSNs was investigated [19] later a more rigorous analysis was conducted in [20]. Then, TinyPK was proposed [21] a scheme which later its vulnerability to the man in the middle attack was shown [22]. Yeh et al. used Elliptic Curve Cryptography (ECC) for user authentication [23]. Security flaw of this method was detected in [24].

To reduce energy consumption of PKC-based systems two different paths in the literature have been pursued, identity-based PKC and hybrid methods. Identity based cryptography differs from the conventional PKC in the sense that authenticity of users' public data do not need explicit verification [25]. First, Hess's identity signature scheme was employed to achieve authenticity of broadcast messages [26]. IMBAS another identity-based method reduced energy consumption of authentication [27]. Later, code dissemination was authenticated based on Rabin-Williams signature [28]. To further reduce energy consumption, Shim *et al.* proposed a pairing-optimal identity-based system with message recovery method [29]. In the second path, mechanisms based on symmetric cryptography are exploited to verify authenticity of users' public data. We adopted name of hybrid for this category. First, Merkle hash tree was exploited [21]. Later, Ren *et al.* employed bloom filter and Merkle hash tree [30]. Another method used ECC and hash functions to authenticate broadcast messages [31]. Finally, in [32] broadcast authenticated PKCs (BA) was proposed. In this method, authentication of public keys was replaced with one-time-signatures based on broadcasted messages from base station (BS).

Energy is one of the main concerns in the WSNs. To improve lifetime of the network more energy efficient mechanisms should be employed. Continuing on our seminal [33], this paper tries to reconcile between high security demand of critical applications and energy consumption of PKC-based systems. This paper makes the following contributions:

- BA method relies on a modified version of μTesla protocol for authenticating messages from BS. Delayed nature of μTesla opens the door for DOS attacks. Such attacks are investigated thoroughly.
- To mitigate DOS attacks, two different strategies based on hash function and bloom filter are proposed.
- Susceptibility of some PKC-based KMSs against battery exhaustion attack is investigated.

The rest of this paper is organized as follows. Section 2 presents BA method and investigates its vulnerabilities against DOS attacks. Section 3 is devoted to the proposed methods. Security analysis and performance of the proposed methods are presented in section 4. Section 5 discusses the proposed methods and finally conclusions are drawn in section 6.

## 2. Basic method and its vulnerability

In Broadcast-Authenticated (BA) method (which we will call the basic method), BS employs message authentication code (MAC), a chain of keys ($\{K_{DSi}\}$), and a set of public-private keys ($P_{ux}$, $P_{rx}$) to generate a set of one-time signatures. Then nodes are preloaded with appropriate credentials and are deployed in the target field. Nodes exchange their signatures and wait for BS to broadcast the corresponding key. Finally, nodes use the exchanged signatures with disclosed key and authenticate each other. Each execution of the protocol is called a cycle and they are repeated according to a timing schedule. According to [33] two different timing schedules are possible. They are depicted in the Figure 1-A. It is noteworthy that, throughout this paper non-uniform timing schedule is employed.

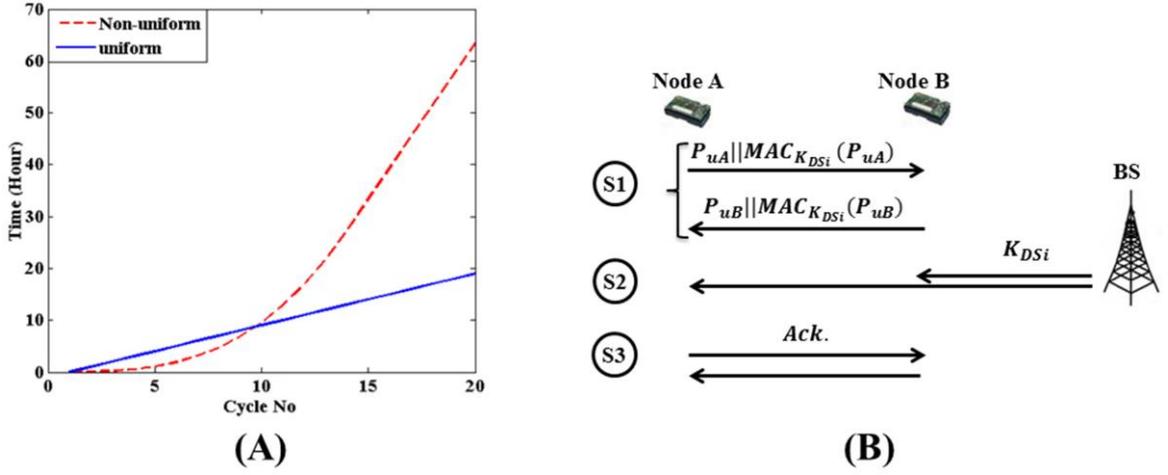

Figure 1  (A) Two possible trimming schedules    (B) Simple schematic of the basic method

Details of the basic method are as follows:

BS generates μTesla key chain. Nodes use these keys for checking authenticity of messages from BS.

$$K_{Authn}\ldots \rightarrow K_{Auth1} \rightarrow K_{Auth0} \rightarrow K_{Auth00} \qquad (1)$$

Then BS generates a key chain for creating a set of one-time signatures.

$$K_{DSn}\ldots \rightarrow K_{DS1} \rightarrow K_{DS0} \qquad (2)$$

For every node, BS generates a pair of public ($P_{ux}$) and private ($P_{rx}$) parameters of elliptic curve Diffie-Hellman (ECDH) and then generates its set of one-time signatures as:

$$\text{Sign}_{x_i} = \text{MAC}_{K_{DSi}}(P_{ux}) \quad i = 1, \ldots, n \qquad (3)$$

If BS can securely convey $K_{DSi}$ to the nodes, they use MAC method for authenticating public key of their neighbours. Finally shared key is calculated according to Diffie-Hellman scheme.  Figure 1-B illustrates main steps of this protocol.

According to Figure 1-B, the basic method has three phases. First, nodes exchange their credentials. Second, BS broadcasts key of those credentials ($K_{DSi}$). Finally, acknowledgment messages are exchanged between nodes. Investigating Figure 1-B shows that, the second step of this method needs a mechanism to guarantee its freshness; otherwise it would be vulnerable against replay attacks. To address this, the basic protocol employed a modified version of μTesla. First, BS encrypts $K_{DSi}$, cycle number, and locally calculated time difference between two consecutive cycles with $K_{Authi}$ and then broadcasts it:

$$BS \rightarrow X: E_{K_{Authi}}(K_{DSi} \| i \| \Delta_i) \qquad (4)$$

After $t$ seconds, BS reveals $K_{Authi}$ for the nodes.

$$BS \rightarrow X: E_{K_{Authi}} \qquad (5)$$

Nodes use $K_{Authi}$ with MAC algorithm and checks authenticity of their neighbours. Then ECDH is invoked for extracting the shared key. Different aspects of this method are discussed in [33]. Although this method is energy efficient, yet authentications of messages from BS are delayed. Next section investigates some possible vulnerabilities of this delay.

## 2.1. Vulnerabilities of the basic method:

### 2.1.1. Network model:

To investigate different scenarios and properties in this paper, different numbers of Mica2 nodes with transceiver range of 30 meters were uniformly distributed over a square field of 500×500 meters. Each simulation was run for 100 times and the final results were averaged. Furthermore, network used packet size of 41 bytes, 32 bytes for payload and 9 bytes for header [34].

### 2.1.2. Adversary model:

In our model, adversary has limited computation power and he cannot break cryptographic primitives. His main interests are to inject false information, authenticate fake nodes to network, and to mount DOS attacks and to exhaust resources of network. To achieve these goals he may eavesdrops communication between nearby nodes and later replay them. Also, he can only compromise limited number of nodes. Furthermore, he may have some agents dispersed throughout network with exclusive frequency band for communication (Figure 4).

### 2.1.3. Flooding attack on the basic method

Delayed authentication, forces nodes to buffer messages and wait for BS to reveal the key. Adversary can take advantage of this delay and flood network with fake messages. If adversary can exhaust memory of nodes, the legitimate message from BS may arrive when there is no memory left. To put this attack into perspective, we assumed that adversary picks his time of attack from a uniform distribution on the interval of $[1, \tau]$. Figure 2-A shows amount of exhausted memory for different values of $\tau$.

In addition to storing bogus messages, nodes retransmit them for their neighbours [35]. This could lead to a severe energy exhaustion attack. To illustrate effect of this attack, we assumed that adversary picks his time of attack from a uniform distribution on the interval [0, 10] minutes. Then, the total amount of energy consumed by all nodes to re-broadcast these fake messages was calculated. Figure 2-B shows how this energy varies with time.

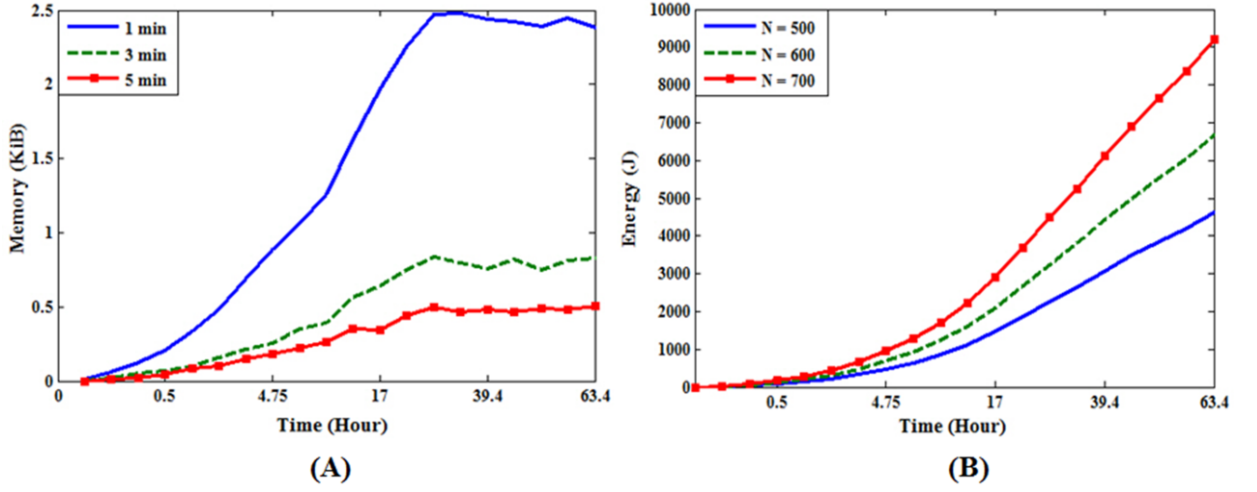

Figure 2  Flooding attacks   (A) Exhausted memory of a node   (B) Exhausted energy of network

## 3. Proposed scheme

In this section we improve the basic method to alleviate its DOS vulnerabilities. We propose two different improved protocols that benefit from immediate authentication. Table I describes notations used in the rest of this paper.

TABLE I. Notations used in this paper

| Notation | Meaning |
|---|---|
| \|\| | Concatenation |
| $P_{ux}$ | Public parameter of elliptic curve Diffie-Hellman of Node $x$ |
| $P_{rx}$ | Private parameter of elliptic curve Diffie-Hellman of Node $x$ |
| $i$ | Cycle number |
| $K_{DSi}$ | Key used to generate $i$th. signature |
| $Tx_i$ | Time measured locally at node $x$ |
| $Sign_{xi}$ | $Sign_{Xi} = MAC_{K_{DSi}}(Pu_X)$ |
| $Ticket_{Xi}$ | $Ticket_{Xi} = P_{uX} \|\| Sign_{Xi}$ |
| $\Delta_i$ | Time difference between two consecutive cycles |
| $K_{Authi}$ | µTesla key chain |
| $MAC_K(M)$ | Message Authentication Code of message ($M$) using key ($K$) |
| $E_K(M)$ | Symmetric encryption of message ($M$) using key ($K$) |
| $K_{AB}$ | Pairwise key between node $A$ and $B$ |
| $f, g$ | Some publicly agreed on functions |
| $h(M)$ | Hash value of message $M$ |
| $\gamma$ | Maximum number of nodes that every node can authenticate in a single cycle |

### 3.1. Broadcast Authenticated protocol with immediate authentication (i-BA)

Equation (4) is concatenation of a key, a counter, cycle duration, all encrypted with a key. If nodes know some information about contents of (4) vulnerabilities of the previous section may be removed. Investigating components of equation (4) shows that BS knows values of counters and both keys ($K_{DSi}$, $K_{Authi}$) for all cycles. We know that BS manages timing of cycles; therefore, assuming that BS (at least) knows time of the next cycle is logical. Consequently, while initiating a new cycle, BS knows message of the next cycle. Thus, BS can append hash value of the next message to the current message. Figure 3 shows a schematic of this concept.

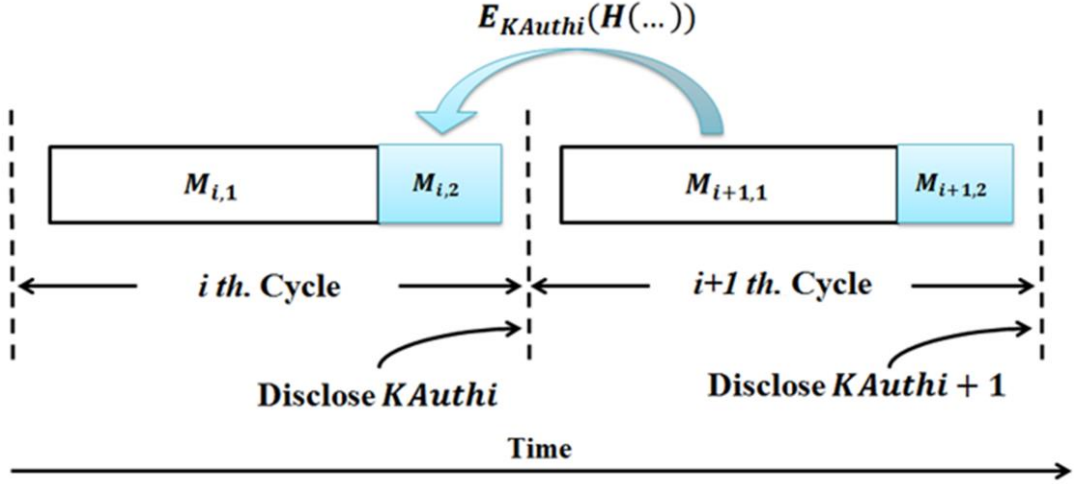

Figure 3 Protocol with Immediate Authentication

Assume node *X* has received following message in the previous cycle:

$$\text{BS} \rightarrow X: E_{K_{Authi-1}}(K_{DSi-1}||i-1||\Delta_{i-1})||E_{K_{Authi-1}}(h(E_{K_{Authi}}(K_{DSi}||i||\Delta_i))||i-1),$$
$$T_{BS} = T_{BSi-1} \qquad (6)$$

Upon receiving of this message nodes have split it into two parts:

$$\begin{cases} M_{i-1,1} = E_{K_{Authi-1}}(K_{DSi-1}||i-1||\Delta_i - 1) \\ M_{i-1,2} = E_{K_{Authi-1}}(h(E_{K_{Authi}}(K_{DSi}||i||\Delta_i))||i-1) \end{cases} \qquad (7)$$

After the key $K_{Authi-1}$ had been disclosed, nodes had decrypted $M_{i-1,2}$ and had saved value of $h(E_{K_{Authi}}(K_{DSi}||i||\Delta_i))$ for the next cycle. For convenience we denote this message by $\mu_{i-1}$. In the next cycle, A receives following message:

$$\text{BS} \rightarrow X: M_{i,1}||M_{i,2} \qquad (8)$$

Fortunately, nodes know hash value of $M_{i,1}$ from the previous cycle and immediately check its authenticity:

$$h(M_{i,1}) \stackrel{?}{=} \mu_{i-1} \qquad (9)$$

After $K_{Authi}$ is disclosed, every node checks its authenticity by performing a hash operation and freshness of message (8) by comparing locally calculated time difference with the one sent from BS. Then, nodes use $K_{DSi}$ with MAC operation to check authenticity of received signatures. Finally, nodes run ECDH and extract shared keys.

The complete *i*-BA protocol is presented in Table II. It is noteworthy that *X* and *Y* denote two neighbouring nodes of network.

TABLE II. *i*-BA PROTOCOL

| |
|---|
| $Ticket_{Xi} = [P_{uX}, Sign_{Xi}], \quad Sign_{Xi} = MAC_{K_{DSi}}(Pu_X)$ |
| $\Delta_i = (T_{BSi} - T_{BSi-1})$ |
| $X \rightarrow Y: Ticket_{Xi}$ |
| $BS \rightarrow X: M_{i,1} \| M_{i,2} \quad , T_{BS} = T_{BSi}$ |
| $X: h(M_{i,1}) \stackrel{?}{=} \mu_{i-1}$ |
| $BS \rightarrow X: K_{Authi} \quad , T_{BS} = T_{BSi} + t$ |
| $X: K_{Authi} \stackrel{h?}{\rightarrow} K_{Authi-1} \; ; T_{xi} - T_{xi-1} \stackrel{?}{\cong} \Delta_i \; ;$ |
| $Sign_{Yi} \stackrel{?}{=} MAC_{K_{DSi}}(P_{u_Y})$ |
| $K_{XY} = f(P_{uY}.P_{rX}, i) = f(P_{uX}.P_{rY}, i)$ |
| $X \rightarrow Y : \mathcal{G}(K_{XY})$ |
| $Y \rightarrow X : \mathcal{G}(K_{XY} + 1)$ |

## 3.2. Bloom filter based Broadcast Authenticated PKC (b-BA)

Bloom filter is a data structure which supports membership queries very efficiently [36]. Bloom filter is an *m*-bit vector all initially set to 0. For representing the set $S = \{s_1, s_2, ..., s_n\}$, *k* independent hash functions are selected such that $h_i(M) \rightarrow [0, m-1], \quad 1 \leq i \leq k$. Then, bits $h_i(s_j), \; 1 \leq i \leq k, 1 \leq j \leq n$ of this vector are set to 1. After bloom filter is constructed, for an element such as *x* if all bits $h_i(x), \; 1 \leq i \leq k$ of bloom filter are equal to 1, then it is said that item *x* belongs to the set *S*.

Considering knowledge of BS about timing schedule, let us go one step further and assume that BS uses a deterministic schedule. Therefore, BS knows all values of $K_{DSi}$, *i*, $\Delta_i$ and can construct the set S={<$K_{DS1}\|1\|\Delta_1$>,<$K_{DS2}\|2\|\Delta_2$>,…,<$K_{DSL}\|L\|\Delta_L$>} where L is the maximum number of authentication cycles. Now, BS can use bloom filter for authenticating its messages. This method is as follows:

First, BS generates signature key chain:

$$K_{DSL} \stackrel{h}{\rightarrow} ... \stackrel{h}{\rightarrow} K_{DS1} \stackrel{h}{\rightarrow} K_{DS0} \stackrel{h}{\rightarrow} K_{DS00} \qquad (10)$$

After constructing the set S, BS constructs its bloom filter. Then, every node is preloaded with its public and private keys, its chain of signatures (3), the last key of key chain ($K_{DS00}$), and bloom filter of S. After nodes are deployed, nodes send their credentials for their neighbours. Then, BS broadcasts message (11):

$$BS \rightarrow X: K_{DSi}\|i\|\Delta_i \qquad (11)$$

Upon receiving this message, every node saves its local time ($T_{Xi}$) and checks authenticity of $K_{DSi}$ by performing a hash function. Then integrity of message (11) is checked with bloom filter. Furthermore, freshness of message (11) is validated by comparing locally calculated time difference with the one sent from BS. If all these security conditions are passed, nodes use $K_{DSi}$ to check validity of received signatures. Finally, nodes run ECDH and extracted shared keys. These steps are shown in Table III.

TABLE III. $b$-BA PROTOCOL

| |
|---|
| $Ticket_{Xi} = [P_{uX}, Sign_{Xi}]$,    $Sign_{Xi} = MAC_{K_{DSi}}(Pu_X)$ |
| $\Delta_i = (T_{BSi} - T_{BSi-1})$ |
| $X \rightarrow Y: Ticket_{Xi}$ |
| $BS \rightarrow X: K_{DSi} \| i \| \Delta_i$ |
| $X: K_{DSi} \xrightarrow{h} K_{DSi-1}$,    $Bloom(K_{DSi}\|i\|\Delta_i) \stackrel{?}{=} 1$, $T_{xi} - T_{xi-1} \stackrel{?}{\cong} \Delta_i$,    $Sign_{Yi} \stackrel{?}{=} MAC_{K_{DSi}}(Pu_Y)$ |
| $K_{XY} = f(P_{uY}.P_{rX}, i) = f(P_{uX}.P_{rY}, i)$ |
| $X \rightarrow Y : g(K_{XY})$ |
| $Y \rightarrow X : g(K_{XY} + 1)$ |

## 4. Analysis of the proposed methods

### 4.1. Security Analysis

#### 4.1.1. Integrity:

This service prevents the unauthorized alteration of data. In the $i$-BA protocol, adversary may try to modify message (6). This message consists of two distinct parts ($M_{i,1}$, $M_{i,2}$) both of which are encrypted with a key that is not disclosed yet. Therefore, their manipulation will produce a random message and it can be detected. $M_{i,2}$ consists of a counter, after decryption nodes use it to check integrity of $M_{i,2}$. Furthermore, this counter chains $M_{i,1}$ and $M_{i,2}$ together, thus preventing attacks like cut and paste [37]. In the $b$-BA protocol, integrity of (11) is the main concern. $b$-BA protocol employs two different mechanisms to provide security. In the first layer, bloom filter provides integrity of (9). Furthermore, messages that pass bloom filter test should have certain values, otherwise they will be discarded. A through discussion on this subject is given in the subsection 4.1.5.

#### 4.1.2. Authenticity:

Authentication is usually divided into two categories. First, data origin authenticity provides assurance about source of message [27]. Both $i$-BA and $b$-BA protocols achieve this by using a hash based key chain. Second, entity authentication addresses identification of different parties of a protocol. Our methods provide this security service by means of one-time signatures of (3).

### 4.1.3. Freshness:

To prevent adversary from replaying old messages, nodes should be able to check freshness of messages. Message (6) and (11) in the *b*-BA and *i*-BA protocol are the main target of replay attacks.

**i-BA protocol:**

Considering the time of replaying message (6), three different scenarios are possible. First, replaying (6) in the same cycle and before disclosure of key $K_{Authi}$. Since message (6) is encrypted with $K_{Authi}$ and this key is not disclosed yet, adversary neither can read its contents nor modify it. Therefore, in this scenario adversary only participates in the blind flooding and distributes message of BS to the nodes. In this scenario, network benefits from replay attack. Second, replaying (6) in another authentication cycle. In another cycle BS reveals $K_{Authl}$. Decryption of an old message (6) with $K_{Authl}$ produces random bits, thus nodes discard such messages. Third, replaying (6) in the same authentication cycle but after $K_{Authi}$ is disclosed. In this scenario, adversary can jam receiver of target node to prevent it from getting message of BS. In this fashion adversary can use $K_{Authi}$ and generate himself valid signature. Then, he can send the forged signature for the target node. Finally, he impersonates BS and discloses $K_{Authi}$ for the target node. Figure 4, depicts this attack.

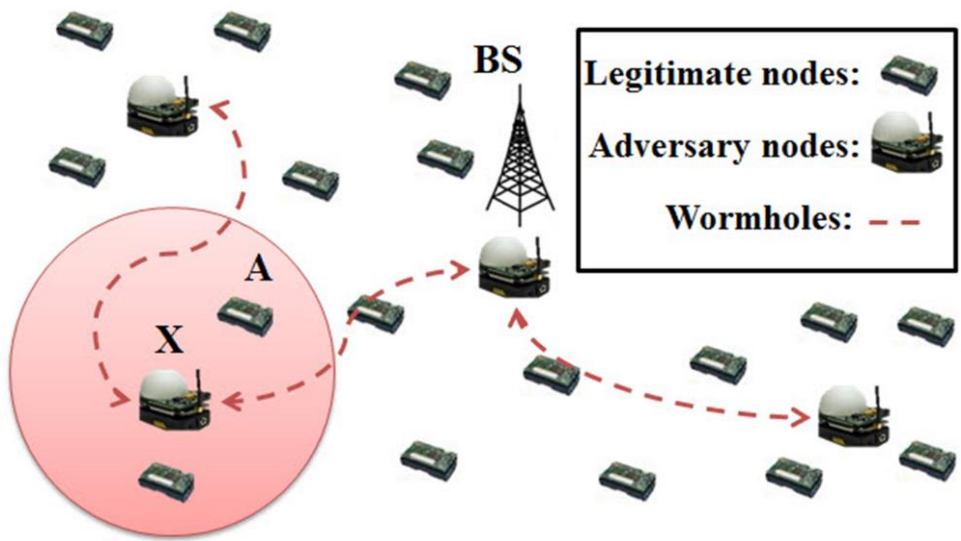

Figure 4 Attacking authentication protocol

According to table II, BS broadcasts message (6) and $K_{Authi}$ at $T_{BSi}$ and $T_{BSi}+t$ respectively. Consequently, adversary replays message (6) at $T_{Xi} > T_{BSi}+t$. According to table II, nodes locally calculate cycle duration and compare it with the one they receive:

$$T_{Ai} - T_{Ai-1} \approx T_{Xi} - T_{BSi-1} \approx \Delta_i + t \quad (12)$$

Equation (12) contradicts security condition of $T_{Ai} - T_{Ai-1} \stackrel{?}{\cong} \Delta_i$ thus this replay attack is detected. Adversary knows value of $K_{Authi}$ and can try to change $\Delta_i$ to $\Delta_i + t$ in (6), but this act violates integrity of (6). We showed that the proposed method detect infringement of integrity, therefore, this attack is also detected

**b-BA protocol:**

In the *b*-BA method two different scenarios are possible. First, replaying (11) in another cycle and second, replaying (11) in the same cycle. These scenarios coincide with the second and the third scenarios of the *i*-BA method, and they are detected.

### 4.1.4. Security and wormhole attack

In this attack two or more malicious nodes collaborate and set up a link with low latency [38]. This can be achieved by using powerful transmitters and another frequency band for exclusive communication. Figure 4 illustrates this attack. In this fashion node *X* gets messages (6) and (11) with negligible delay. Consequently, he can use disclosed key and generate valid signature without changing value of $\Delta_i$.

Let us investigate authentication cycles. Cycle *i* starts with receiving of $K_{Authi-1}$ and $K_{DSi-1}$ and lasts until $K_{Authi}$ and $K_{DSi}$ are received, in the *i*-BA and *b*-BA protocol respectively. If nodes terminate each cycle *t* seconds before receiving of $K_{Authi}$ and $K_{DSi}$ this attack is prevented. In this fashion when adversary generates his signatures, the cycle is terminated and thus nodes will discard his forged signatures.

### 4.1.5. False positive value of Bloom filter

Employing bloom filter introduces a false positive into the scheme. It means that bloom filter may suggest that an element *x* is in *S*, even though it is not. According to [39] the lowest value of this probability is equal to $2^{-k}$ and it is achieved for $k=(m.ln2)/n$, where *k*, *m,* and *n* represent number of hash functions, length of bloom filter, and cardinality of set *S*. We want to estimate probability of forging message (11). Security of *b*-BA relies on two layers. In the first layer, message should pass bloom filter. A randomly generated message passes bloom filter with the probability of $2^{-ln2.(m/n)}$. In the second layer, contents of the message are checked. Message (11) consists of three components, a key ($K_{DSi}$), a cycle counter (*i*), and a time difference ($\Delta_i$). Two of these components are deterministic and they should satisfy:

$$\underbrace{K_{DSi} \xrightarrow{h} ... \xrightarrow{h} K_{DS0} \xrightarrow{h} K_{DS00}}_{i+1 \text{ times}} \quad (13)$$

Let $L_k, L_i$, and $L_\Delta$ denote length of $K_{DSi}$, *i*, and $\Delta_i$ in bits. Probability of a random message to pass second security layer is equal to:

$$\frac{2^{L_\Delta}}{2^{L_\Delta + L_i + L_k}} = 2^{-(L_i + L_k)} \quad (14)$$

Consequently, probability of forgery is equal to:

$$2^{-(L_i+L_k)} \times 2^{-\ln 2 \cdot (m/n)} \qquad (15)$$

### 4.2. Connectivity of the proposed methods:

Both methods assume that all nodes receive BS messages. Apparently If BS is equipped with a powerful transmitter this assumption is correct. But if its range is limited, nodes can retransmit messages of BS for their neighbours. In this fashion BS messages can propagate through network.

*Theorem1:* Let $p_{loss}$ and $k$ denote probability of packet loss and number of neighbours of node *C*, then probability of *C* receiving message of BS satisfies:

$$p_r = 1 - p_{loss}^{k.p_r} \qquad (16)$$

*Proof*: Among neighbours of node *C*, on average $k.p_r$ of them have received message of BS. Thus, probability of node *C* not receiving message of BS is:

$$p_{fail} = p_{loss}^{k.p_r} \qquad (17)$$

*Theorem2:* If two nodes participate in *m* authentication cycles, then probabilities of sharing a key in the *i*-BA and *b*-BA method are equal to:

$$P_{mi-BA} = 1 - (1 - p_r^4)^m \qquad (18)$$
$$P_{mb-BA} = 1 - (1 - p_r^2)^m \qquad (19)$$

*Proof*: If nodes receive message of BS with probability of $p_r$ and λ denotes number of BS messages that node A should receive to run the protocol (λ is equal to 2 and 1 in the *i*-BA and *b*-BA methods), then probability of both nodes A and B receiving necessary messages of BS is:

$$P_{success} = p_r^\lambda \cdot p_r^\lambda = p_r^{2\lambda} \qquad (20)$$

If nodes A and B participate in *m* authentication cycles, probability of sharing a key would be equal to:

$$P_m = 1 - (1 - P_{success})^m \qquad (21)$$

Figure 5 shows how these probabilities change.

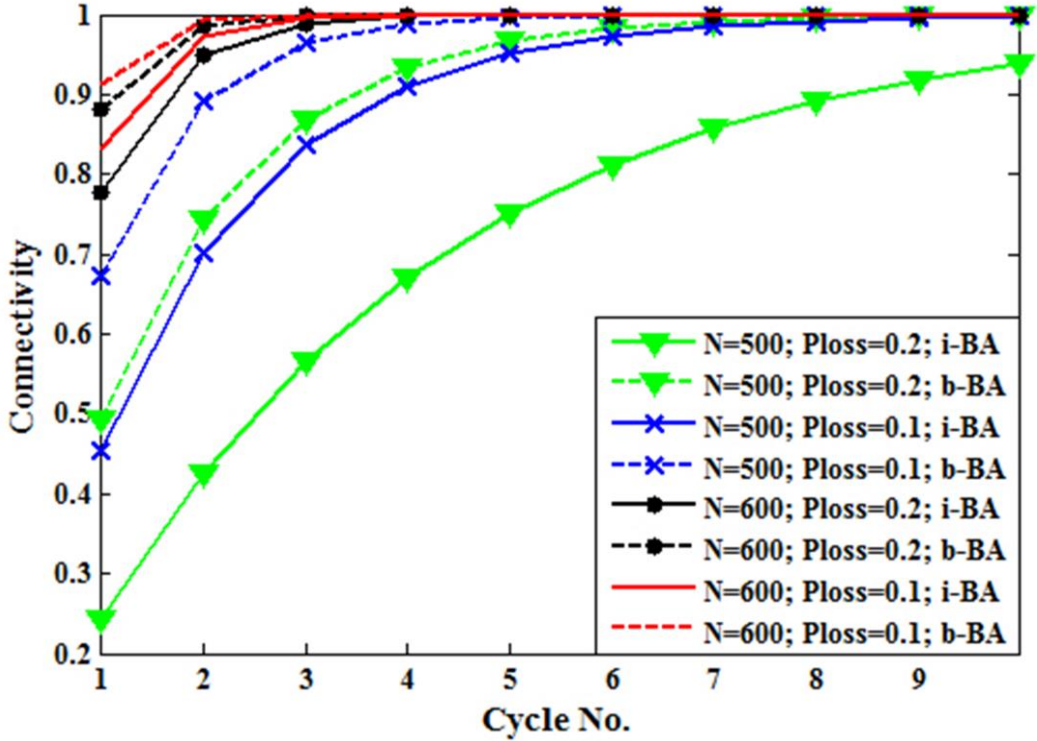

Figure 5 Probability of sharing a key after participating in *m* authentication cycles.

### 4.3. Energy consumption

To calculate energy consumption, cost of transmission [34] and executing cryptographic primitives [34, 40] were added together. Parameters of this calculation were as follows: $L_i = 10$ bits, $L_\Delta = 14$ bits, 128 bits for MAC and all keys, and 160 bits for ECDH keys. Furthermore parameters of bloom filter were: $m=2^{15}$, $k=23$, $n = 2^{L_i} = 1024$. Finally, SHA-1 and AES methods were used for symmetric encryptions and hash operations. Adding communication and computation costs of table II and table III, total cost of *i*-BA and *b*-BA became 60.50 mj and 62.54 mj, respectively.

### 4.4. Resiliency against Battery Exhaustion Attack

Energy is the most precious resource in WSNs. Thus, numerous attacks have aimed to exhaust it. The main purpose of these attacks is to force victims to run costly operations. This section investigates resiliency of some PKC-based KMSs to these attacks. To mount an effective attack, adversary may listen to ongoing traffics ($Ticket_{Xi}$) and later re-transmits them for the victims. Such data will be authenticated and the whole protocol will be executed, therefore this scenario will lead to the the most severe battery exhaustion attack. We assumed that adversary picks his time of attack according to a uniform distribution on the interval of [0, 10] minutes. Furthermore, the maximum number of authentication in a single cycle was limited to 8 ($\gamma=8$). Figure 6 shows exhausted energy of a single node for different PKC-based KMS in the logarithmic scale.

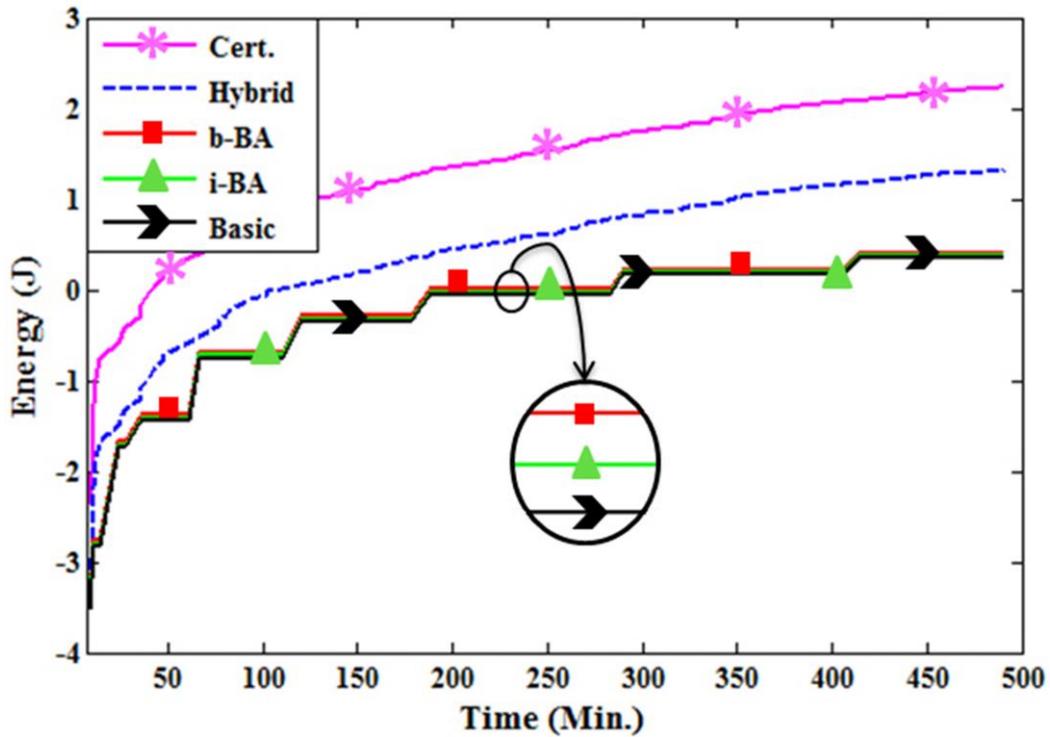

Figure 6   Exhausted energy of a single node. Notations are in accordance with those of Table IV.

### 4.5. Scalability

Let us assume that ID of nodes is 2 bytes and IDs of revoked nodes should be stored. Assuming 64KiB of memory, it is possible to calculate maximum size of network in each method. $b$-BA uses $2^{15}$ bits of memory for storing bloom filter, therefore it supports $(2^{19}-2^{15})/16=30720$ nodes. On the other hand, $i$-BA method uses all this memory for revoked nodes therefore it can support up to $2^{19}/16=32678$ nodes.

### 5. Discussion

Nodes of WSNs are resource constraint and they are prone to different types of attacks. Thus, their protocols should address these problems. PKC-based KMS have many desirable characteristics. They provide perfect resiliency, perfect global, local, and node connectivity, and they have very good scalability and extensibility properties. But their energy consumption is a heavy burden on tight resources of nodes. Therefore, reducing energy consumption of PKC-based KMS is very desirable. A possible solution is to replace checking of digital certificate with symmetric based systems. One such method relied on a set of one-time signatures. For this method to work, BS should convey some credentials to nodes securely. To this end, a modified version of μTesla was proposed in the basic method. According to figure 2, this system is vulnerable to flooding attacks. Furthermore, these figures show that, effect of this DOS attack on the energy consumption is more severe.

To alleviate these problems we assumed a more knowledgeable BS scenario. First, we assumed that BS knows exact time of the next authentication cycle. In this manner, hash value of the next message was appended to the current message. Second scheme, went one step further and assumed that BS knows time of all authentication cycles. This knowledge enabled BS to use bloom for authentication. Both of these assumptions added immediate authentication to the system and solve its vulnerability to flooding attacks.

While the proposed methods improved resiliency of KMS against DOS attacks, the proposed methods inherits other desirable characteristics of the basic method. One of those properties is addressing problem of dead nodes. Because battery of dead nodes has been depleted they have lost their functionality. But they have very valuable information stored in them. Adversary can collect these nodes and exploit their information for mounting more effective attacks. For example, he can read their keying material and use them for programming his nodes. Therefore, KMS should employ a reliable revocation mechanism. One solution is to broadcast ID of revoked nodes. Apparently, nodes have limited amount of memory and storing information of many dead nodes would be infeasible. Another solution that is very common to PKC-based system is to assign an expiration time to each certificate. In the proposed methods problem of dead nodes can effectively be addressed. The proposed methods rely on one-time signatures. Also, as BS reveals key of signatures, those signatures get expired. Considering average life-time of nodes BS can preload nodes with suitable number of signatures such that when their battery is depleted, there would no valid signature be left. Consequently, dead nodes would not provide useful information for adversary.

Comparing results of previous sections show that both $i$-BA and $b$-BA have their own merits. $i$-BA consumes less energy and it supports larger networks, on the other hand in the $b$-BA scheme nodes share the common key much faster, especially when network in not dense or probability of packet loss is high (figure 5). Furthermore, $b$-BA method is simpler and its implementation would be easier. Finally, according to figure 6, both of the proposed methods have good resiliency against battery exhaustion attack. Table IV presents a comparison between proposed methods and some previous PKC-based KMSs.

TABLE IV. Comparison between different PKC based key management systems

| Scheme | Abbreviation | Energy cost | Scalability | Flooding Attack | Battery Exhaustion | Reference |
|---|---|---|---|---|---|---|
| Certificate Based | Cert | 187.6 | Highest | Resilient | Very Low Resiliency | [34] |
| Hybrid Method | Hybrid | 75.26 | Moderate | Resilient | Low Resiliency | [26] |
| Basic Method | Basic | 58.68 | Highest | Vulnerable | High Resiliency | [32] |
| $i$-BA | $i$-BA | 60.50 | Highest | Resilient | High Resiliency | ------ |
| $b$-BA | $b$-BA | 62.54 | Very High | Resilient | High Resiliency | ------ |

## 6. CONCLUSION

There are lots of new challenges in WSNs due to different trade-offs and conflicting requirements. Tight constraints on resources -such as energy, memory and computation power- in addition to unique features of these networks, have turned most of algorithms for the conventional network impractical. Recently, security and KMS in WSNs have received lots of attention. Unfortunately, most of previous works have sacrificed security in favour of reducing energy consumption. To avoid such trade-offs, one solution is to reduce energy consumption of PKC-based KMS. Pursuing this goal, this paper investigated an energy efficient KMS. It was shown that delayed authentication of this method lead to some serious DOS attacks. Later, we showed that by extending knowledge of BS these vulnerabilities can be solved. To this end, two new methods based on adding hash value of the next message to the current message and bloom filter were proposed. Simulation results showed that these methods maintain energy efficiency, high scalability, and high resiliency against battery exhaustion attack of the basic method. Furthermore, immediate authentication of these improved methods removed vulnerability of the basic method to flooding attacks.